\documentclass[11pt]{article}
\usepackage{amssymb,amsthm,amsmath,float,epsfig}
\usepackage[small,width=0.95\linewidth]{caption}
\usepackage[total={6.5in,9.0in}, top=1.0in, left=1.0in, includefoot]{geometry}
\usepackage{graphicx}
\usepackage{enumerate}
\usepackage{enumitem}
\usepackage{bm}
\usepackage{authblk}
\usepackage{setspace}
\newcommand{\appropto}{\mathrel{\vcenter{
  \offinterlineskip\halign{\hfil$##$\cr
    \propto\cr\noalign{\kern2pt}\sim\cr\noalign{\kern-2pt}}}}}

\graphicspath{{figures/}}

\title{Evolving network structure of academic institutions}
\author{Shufan Wang}
\author{Mariam Avagyan}
\author{Per Sebastian Skardal}
\affil{Department of Mathematics, Trinity College, 300 Summit Street, Hartford, CT 06106, USA}
\date{}

\begin{document}

\maketitle

\begin{abstract}
Today's colleges and universities consist of highly complex structures that dictate interactions between the administration, faculty, and student body. These structures can play a role in dictating the efficiency of policy enacted by the administration and determine the effect that curriculum changes in one department have on other departments. Despite the fact that the features of these complex structures have a strong impact on the institutions, they remain by-and-large unknown in many cases. In this paper we study the academic structure of our home institution of Trinity College in Hartford, CT using the major and minor patterns between graduating students to build a temporal multiplex network describing the interactions between different departments. Using recent network science techniques developed for such temporal networks we identify the evolving community structures that organize departments' interactions, as well as quantify the interdisciplinary centrality of each department. We implement this framework for Trinity College, finding practical insights and applications, but also present it as a general framework for colleges and universities to better understand their own structural makeup in order to better inform academic and administrative policy.
\end{abstract}

\section*{Introduction}
The organizational structures of today's higher education academic institutions are exceedingly complex with few exceptions~\cite{berger2002influence}. In particular, modern colleges and universities are comprised of many different departments that interact with one another through various faculty and student activity~\cite{hillier1991visible}. Additionally, most universities are organized into multiple schools and virtually all colleges and universities offer unique programs and concentrations that facilitate further inter-department interactions~\cite{toma2010building}. Unsurprisingly, the complex structures of these colleges and universities have a significant impact on both the scientific and scholarly production of their faculty members and the academic, social, and eventually professional endeavors of their students~\cite{pascarella2005college,pascarella2006college}. Previous studies have investigated the structure of interactions between {\it different} colleges and universities via hiring networks~\cite{clauset2015systematic} and the scientific co-authorship~\cite{newman2006finding}, and social and communication networks have been studied {\it within} individual institutions~\cite{bernard1980informant,guimera2003self}. However, little is known about the structure of the {\it academic interactions} within a given college or university. Instead, institutions are led to make significant decisions and craft policy based on simplistic statistics such as class size and the distribution of degrees awarded by various departments, which ignores key information such as which departments are more similar or strongly linked to other departments according to more specific metrics. Therefore, there is a need to extract and interpret more nuanced information characterizing the structural patterns in our colleges and universities in order to develop more efficient policies to better serve the faculty and the student body alike.

The study of networks has emerged as a uniquely fruitful area of research, yielding important theoretical tools for understanding real-world complex structures and systems~\cite{strogatz2001exploring}. At its structural core a network is mathematically represented by a graph -- a collection of nodes and the edges connecting them~\cite{newman2003structure,boccaletti2006complex}. Applications of network science approaches are widespread, ranging from understanding microscopic phenomena, e.g., protein-protein interactions~\cite{han2004evidence,vazquez2003global} and gene-regulation~\cite{teichmann2004gene,huang2005cell}, to macroscopic phenomena, e.g., social interactions between people~\cite{palla2005uncovering,bagrow2012mesoscopic} and large-scale power grids~\cite{rohden2012self,motter2013spontaneous}. Two concepts that are particularly useful for understanding the structural patterns of a network are community structure and centrality. Community structure refers to partitions of a network into groups such that many links connect nodes within the same group, but few links connect nodes in different groups~\cite{newman2012communities}. Thus, the community structure of a network describes a natural organization of the network into groups of closely-related nodes as defined by the network structure. On the other hand, centrality refers to a measure of the standing of each individual node in a network compared to others~\cite{wasserman1994social}. Therefore, centrality measures are useful in identifying individual nodes that are important for connecting the overall network. While many different network centralities exist, each in some way describes the importance of each node in terms of connecting the overall network.

In this paper we study the network structure of our home institution, Trinity College, in Hartford, CT. We begin by constructing a network describing the academic interactions between different departments using the major and minor patterns of graduating seniors. This academic network is time-varying, and thus is a natural example of a temporal multiplex network. Using techniques recently developed for such temporal multiplex networks we study the community structure of Trinity College and the interdisciplinary centrality of its various departments as they evolve through the years. Our results shed a great deal of light on the structural patterns of Trinity College, offering practical insights into the structure of the institution that we believe might better inform the creation and implementation of policy. For instance, several communities exist that highlight important groups of well-connected departments. Interestingly, the communities that emerge differ from the typical science vs humanities separation that one might expect -- instead we find that the community structure is much more nuanced. This suggests a less unified academic environment than might be ideal and the possibility for various policies to affect departments in various communities much differently. Moreover, we identify certain ``stalwart'' departments that remain in the same community through the years, while other departments are more flexible in their standing, belonging to multiple communities as years pass. We also use network centrality techniques to identify those departments that are particularly important in terms of connecting the whole environment. Interestingly, departments that are more central do not necessarily correspond to those departments that are larger. We also identify departments that act as strong connectors due to their majors, while other departments act as strong connectors due to their minors. Finally, we close with a discussion of our results and an outlook into their possible applications and use at other institutions.

\section*{The Network}
We start by describing the construction of the academic network of Trinity College. To begin, we identify the full range of departments at the college that offer all possible major and minor degrees. At Trinity College we identify 32 such departments, assigning each one a distinct four-letter code, which is summarized in Table~\ref{tab1} in the Appendix. For example, the anthropology and mathematics departments are represented ANTH and MATH, respectively. Next, for each graduating year we identify all students that earned a degree from two or more departments. Each of these students then contributes to one or more interactions between different departments of the form major-major, major-minor, or minor-minor. For instance, a student that completes a double-major in engineering and mathematics contributes one major-major interaction between engineering and mathematics. On the other hand, a student that completes a major in English with minors in sociology and film contributes three interactions: two major-minor interactions between english and sociology and english and film, respectively, and one minor-minor interaction between sociology and film. For each of the three types of interactions we create an adjacency matrix that represents the number of interactions for the class graduating in year $t$, denoting them $\tilde{A}_{\text{maj-maj}}^{(t)}$, $\tilde{A}_{\text{maj-min}}^{(t)}$, and $\tilde{A}_{\text{min-min}}^{(t)}$. Since our college consists of 32 departments, each adjacency matrix is 32 by 32, representing the interactions between 32 nodes. Each interaction contributes a link of weight one to the corresponding entry in the appropriate matrix. Finally, we interpret each interaction as undirected, so that each resulting adjacency matrix is symmetric.

Using the process outlined above we obtain for each graduating class three adjacency matrices, each describing the relationships between departments via major-majors, major-minors, and minor-minors. In order to combine these topologies into one overall network we introduce a parameter $\alpha\in[0,1]$ describing the relative importance of a minor in comparison to a major. Specifically, for the class graduating in year $t$ we build the overall adjacency matrix $\tilde{A}_\alpha^t$, defined as
\begin{align}
\tilde{A}_\alpha^{(t)}=\tilde{A}_{\text{maj-maj}}^{(t)}+\alpha \tilde{A}_{\text{maj-min}}^{(t)}+\alpha^2 \tilde{A}_{\text{min-min}}^{(t)},\label{eq:A1}
\end{align}
such that, in comparison to major-major interactions, major-minor and minor-minor interactions are weighted by a fraction of $\alpha$ and $\alpha^2$, respectively. In principle one can choose $\alpha$ to be the average number of courses required for a minor as compared to a major, or one can vary $\alpha$ to examine the effect that minors play in altering the structure of the network (as we do below). Finally, we note that makeup of the student body any given year consists not only of the students that graduate that year, but also those that eventually graduate in each of the three following years. Therefore, the adjacency matrix we use to describe the network of the academic environment in year $t$, denoted $A_\alpha^{(t)}$, is created by combining the adjacency matrices of the graduating classes for years $t$, $t+1$, $t+2$, and $t+3$:
\begin{align}
A_\alpha^{(t)}=\tilde{A}_\alpha^{(t)}+\tilde{A}_\alpha^{(t+1)}+\tilde{A}_\alpha^{(t+2)}+\tilde{A}_\alpha^{(t+3)}.\label{eq:A2}
\end{align}

\begin{figure}[t]
\centering
\epsfig{file =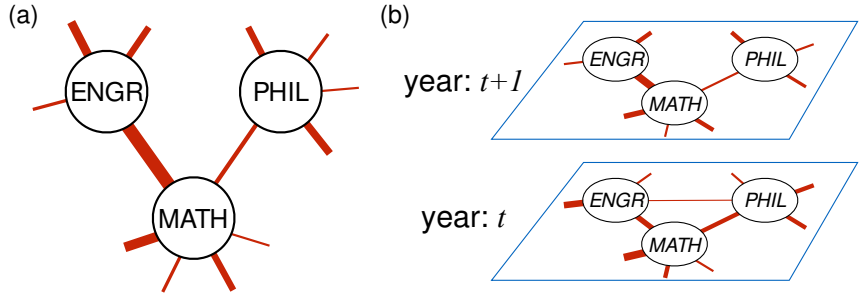, clip =,width=0.7\linewidth }
\caption{{\bf Network structure example.} Illustration of a small portion of the network for (a) a single year and (b) a multiplex network constructed from two adjacent years.} \label{fig1}
\end{figure}

The adjacency matrix $A_\alpha^{(t)}$ thus describes the interaction between different departments via their students' major and minoring patterns in a year $t$ for a chosen minor parameter $\alpha$. In principle $A_\alpha^{(t)}$ is weighted but remains symmetric. A small portion of such a network, describing the engineering (ENGR), mathematics (MATH), and Philosophy (PHIL) departments, is illustrated in Fig.~\ref{fig1}(a). Note that the network is both undirected and weighted. Such a network can be obtained for a range of several years, giving rise to a multi-layered temporal network where the network for each distinct year comprises a different layer, as illustrated in Fig.~\ref{fig1}(b). Note that each different layer, corresponding to the network at a different year, consists of the same collection of nodes, but with different connection patterns, thus contributing to a temporal multiplex network~\cite{de2013mathematical,kivela2014multilayer}. We also note that, since the most recent year available in our dataset is 2016, the most recent layer in our network is that for 2013 (which includes the graduating classes 2013, 2014, 2015, and 2016). In the remainder of this paper we investigate the structural features of this temporal multiplex network representing the academic interactions at Trinity College, first focusing on community structure, then on centrality.

\section*{Community Structure}
Many real world networks display a key feature known as community, or modular, structure: a partition of the nodes into two or more groups where nodes share many links with nodes in the same group, but few with nodes in other groups, relatively speaking~\cite{newman2012communities}. The identification of communities thus provides a valuable description of the structure of the network and has many different applications in many different contexts such as groups of friends in social networks and similar species in food webs~\cite{girvan2002community}. In the case of the academic network studied here, the identification of communities not only allows us to better understand the network structure, but has more specific utility. For instance, knowledge of the community structure might allow institutions to better predict what groups of departments may be more or less impacted by certain policies or understand what other departments will be more or less affected by a curriculum change in another department. Denoting the community to which node $i$ belongs in a given partition as $s_i$, the community structure of a single-layer network represented by the adjacency matrix $A$ can be identified by maximizing the modularity $Q$~\cite{newman2004finding}, defined as
\begin{align}
Q = \frac{1}{N\langle k\rangle}\sum_{ij}\left(A_{ij}-\frac{k_ik_j}{N\langle k\rangle}\right)\delta(s_i,s_j),\label{eq:Q1}
\end{align}
where $k_i=\sum_{j}A_{ij}$ is the (possibly weighted) degree of node $i$, $\langle k\rangle=N^{-1}\sum_{i}k_i$ is the mean degree, and $\delta(s_i,s_j)$ is the Kronecker delta function that evaluates to one if $s_i=s_j$ and zero otherwise. In practice, finding community structure in large networks is a difficult problem, however several methods exist for identifying community structures including aggregative~\cite{clauset2004finding}, divisive~\cite{duch2005community}, and spectral~\cite{newman2006finding} methods. Here we use the divisive method of extremal optimization~\cite{duch2005community}.

\begin{figure}[t]
\centering
\epsfig{file =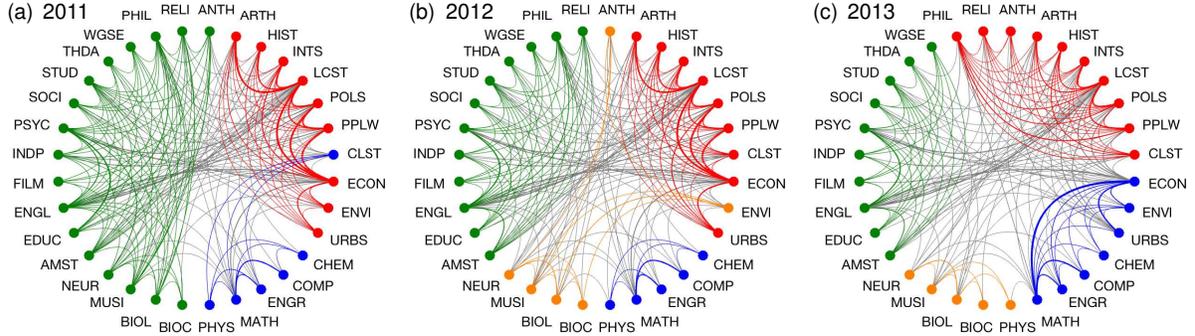, clip =,width=0.96\linewidth }
\caption{{\bf Community structure: single layers.} Community structure, indicated by color, for the networks from the three most recent years of 2011, 2012, and 2013. Minor parameter: $\alpha=0.5$.} \label{fig2}
\end{figure}

We begin by studying community structure in single layers of the network, constructed using a minor parameter of $\alpha=0.5$, corresponding to a weighting where major-minor interactions are half as significant and double major interactions and double minor interactions are a quarter as significant. In principle one could estimate $\alpha$ as the ratio of the average number of credits required for institution-wide minors to the average number of credits required for institution-wide majors. Here we make this simple choice, noting that we will consider varying $\alpha$ below. In Fig.~\ref{fig2} we illustrate community structure found in the single-layer networks from the three most recent years of 2011, 2012, and 2013, indicating different communities by color. Departments are presented in an order that best groups departments in the same community, and in the same order through the three years. In 2011 we identify three communities, roughly corresponding to historical humanities (red, e.g., economics, history, and political science), artistic humanities and descriptive sciences (green, e.g., english, religion, biology, and neuroscience), and finally the quantitative science (blue, e.g., engineering, mathematics, and physics). Note that this partition into communities is significantly different from the sciences vs. humanities separation that one may expect. In particular, while the quantitative sciences constitute a community, the descriptive sciences belong to the same community as the artistic humanities. In 2012 we observe a significant change via the birth of a new community, roughly corresponding the descriptive sciences (orange). This community is primarily made up of departments which belonged to the artistic humanities the previous year, but also includes anthropology and environmental science, both of which belonged to the historical humanities. Also, classical studies department switched from the quantitative sciences community to the historical humanities community. Finally, more changes are observed in 2013: physics joins the descriptive science community, economics, environmental science, and urban studies join the quantitative science community, and philosophy and religion join the historical humanities.

The year-to-year variation in the communities described above indicates the need for a more nuanced approach for understanding the evolution of community structure through time~\cite{mucha2010community}. In particular, while the overall composition of communities from year-to-year share similar properties, we observe both the split of one community into two as well as switching of some department from one community to another. A natural question then arises: do we still observe such phenomena if a given node's community membership in two adjacent layers is connected? In order to answer this question we turn to recent work where the concept of modularity has been formulated for the case of temporal multiplex networks~\cite{bazzi2016community}. In particular, we now designate the community of node $i$ in each layer $t$ by $s_i^{(t)}$, and adopt the multilayer modularity formulation
\begin{align}
Q_{\omega}&=\frac{1}{L}\sum_{t=1}^L\left[\frac{1}{N\langle k^{(t)}\rangle}\sum_{i,j=1}^N\left(A_{ij}^{(t)}-\frac{k_i^{(t)}k_j^{(t)}}{N\langle k^{(t)}\rangle}\right)\delta\left(s_i^{(t)},s_j^{(t)}\right)\right]\\&+\frac{2\omega_{\text{mod}}}{N(L-1)}\sum_{t=1}^{L-1}\sum_{i=1}^N\delta\left(s_i^{(t)},s_i^{(t+1)}\right),\label{eq:Q2}
\end{align}
where $L$ is the total number of layers in the multiplex, $k_i^{(t)}$ is the degree of node $i$ in layer $t$, and $\langle k^{(t)}\rangle$ is the mean degree in layer $t$. We note that the formulation of the multilayer modularity in Eq.~(\ref{eq:Q2}) has two contributing terms and is a slight modification (up to a rescaling of $Q_{\omega}$ and $\omega_{\text{mod}}$) of that in Ref.~\cite{bazzi2016community}. The first term accounts for the modularity within each individual layer and the second term, which includes a {\it modularity persistence} parameter $\omega_{\text{mod}}>0$ accounts for the agreement in the communities for the same node in two adjacent layers. Thus, $\omega_{\text{mod}}$ modifies the degree to which the communities of the same node in subsequent layers is preferred to be the same, i.e., persist. In the limit $\omega_{\text{mod}}\to0^+$ persistence has no effect on the multilayer modularity and the resulting community structure is simply that of each individual layer, while larger values of $\omega_{\text{mod}}$ dictate a preference for nodes in adjacent layers to remain in the same community, thereby unifying the community structure of the multiplex.

\begin{figure}[t]
\centering
\epsfig{file =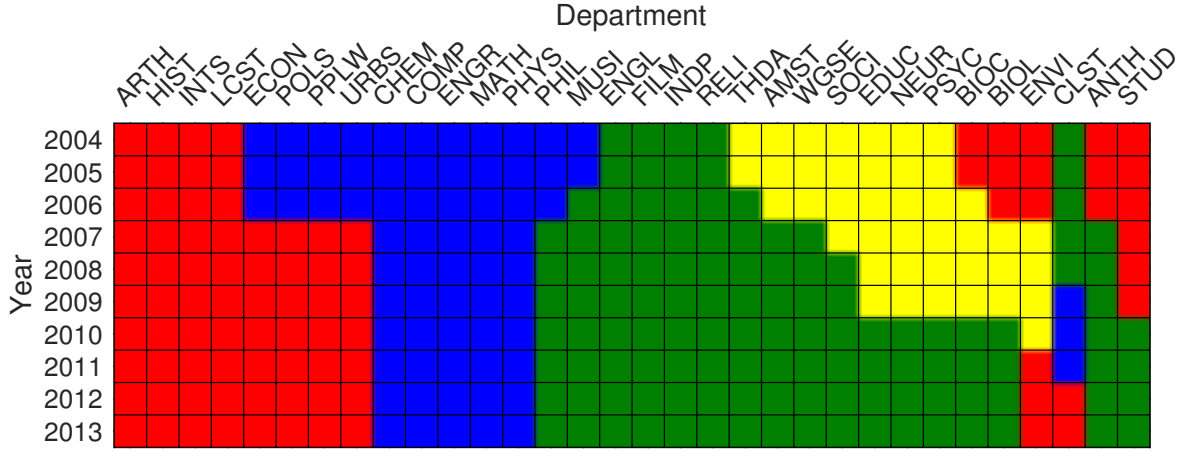, clip =,width=0.95\linewidth }
\caption{{\bf Community structure: temporal network.} Evolution of community structure, indicated by color, throughout the college from 2004-2013. Minor parameter: $\alpha=0.5$. Persistence parameter: $\omega=0.2$.} \label{fig3}
\end{figure}

In Fig.~\ref{fig3} we illustrate the community structure, indicated by color, found in the 10-layered multiplex consisting of the years 2004--2013 for a minor parameter value $\alpha=0.5$ and persistence value of $\omega_{\text{mod}}=0.2$, which we find nicely balances the effects of persistence vs. modularity in individual layers. (Community structure is found using a modification of the extremal optimization technique and is summarized in the Appendix.) Departments are presented in an order that best groups departments in the same communities, easing the visual identification of different communities and their evolution. Overall, we observe four different communities and a complex pattern of structure. First, several ``stalwart'' departments exist that remain in the same community over all ten years: art history, history, international studies, and language and culture studies form a backbone of the historical humanities community (red), chemistry, computer science, engineering, mathematics, and physics form the backbone of the quantitative science community (blue), and english, film studies, and religion (as well as the individualized degree program) form the backbone of the artistic humanities community (green). We note that physics belonged to different communities when considering layers in isolation [see Fig.~\ref{fig2}], but with the added preference for agreement between nodes in adjacent layers via the persistence parameter $\omega_{\text{mod}}$, physics becomes a stalwart of the quantitative sciences. Another community also exists (yellow) comprised of the descriptive sciences and some other humanities, but is extinguished by the year 2011, by which time most of its members have joined the artistic humanities community. Again, the effect of persistence is observed in the descriptive sciences: in the single layers for 2012 and 2013 the descriptive sciences comprised its own community [see Fig.~\ref{fig2}], but this is not true for the overall multiplex with $\omega_{\text{mod}}=0.2$. Rather, the effect of persistence is to keep the descriptive sciences in the same community as the artistic humanities. Finally, we observe in many instances multiple departments switching communities simultaneously or approximately at the same time. For instance, economics, political science, public policy and law, and urban studies all switch from the quantitative science community to the historical science community at the end of 2006. Moreover, philosophy, music, theater and dance, American studies, women, gender, and sexuality, sociology, and anthropology all join the artistic humanities between 2005 and 2008.

\section*{Centrality}
As a complement to the features of our academic network captured by community structure, we also study the centrality properties of our network. While a great many centrality measures exist for a given network, each with slightly different meanings, all centralities measure in some sense each node's role or importance in connecting the network~\cite{newman2003structure}. Moreover, many of the most useful centrality measure are represented by eigenvectors of a matrix, for instance PageRank centrality~\cite{gleich2015pagerank}, hub and authority centrality~\cite{kleinberg1999authoritative}, dynamical importance~\cite{restrepo2006characterizing}, and classical eigenvector centrality~\cite{newman2003structure}. For a single-layered network any eigenvector-based centrality measure is described by the the dominant eigenvector of a matrix $C$ that is some function of the adjacency matrix $A$~\cite{maccluer2000many}. For instance, in the case of PageRank centrality the centrality $c_i$ of node $i$ is given by $v_i$ where $\bm{v}$ is the leading eigenvector of the matrix $C=(D^{\text{in}})^{-1}A$, where $D^{\text{in}}=\text{diag}(k_1^{\text{in}},\dots,k_N^{\text{in}})$.

Recently Taylor et al.~\cite{taylor2015eigenvector} formulated the centrality problem for a temporal multiplex network for any eigenvector-based centrality. Given a temporal multiplex as we study here with adjacency matrices $A^{(1)},\dots,A^{(L)}$ for the different layers the centrality matrices $C^{(1)},\dots,C^{(L)}$ are computed and used to construct the supra-centrality matrix
\begin{align}
\mathbb{C}=\begin{bmatrix} C^{(1)} & \omega_{\text{cen}}I & 0 & \cdots & 0 \\
\omega_{\text{cen}}I & C^{(2)} & \ddots &  & \vdots \\
0 & \ddots & \ddots & \ddots &0 \\
\vdots & & \ddots & C^{(L-1)} & \omega_{\text{cen}}I \\
0 & \cdots & 0 & \omega_{\text{cen}}I & C^{(L)}
\end{bmatrix},\label{eq:C}
\end{align}
where $\omega_{\text{cen}}$ represents a centrality persistence measure with a similar interpretation as the modularity persistence parameter used above. The centrality of each node in each layer is then given by the dominant eigenvector of $\mathbb{C}$, which comes in the form
\begin{align}
\bm{v} = [\bm{v}^{(1)T}|\bm{v}^{(2)T}|\cdots|\bm{v}^{(L)T}]^T.\label{eq:v}
\end{align}
Finally, since the centrality of a given node may differ significantly from layer-to-layer, i.e., the values of $\bm{v}^{(t)}$ may differ on average significantly from those of $\bm{v}^{(t')}$ we compute the conditional centralities of each node in each layer, defined as
\begin{align}
u_i^{(t)}=\frac{v_i^{(t)}}{\sum_{j=1}^Nv_j^{(t)}}.\label{eq:u}
\end{align}
Specifically, the conditional centralities normalize the centralities in each layer to one, quenching any effect of layer-to-layer effects.

\begin{figure}[t]
\centering
\epsfig{file =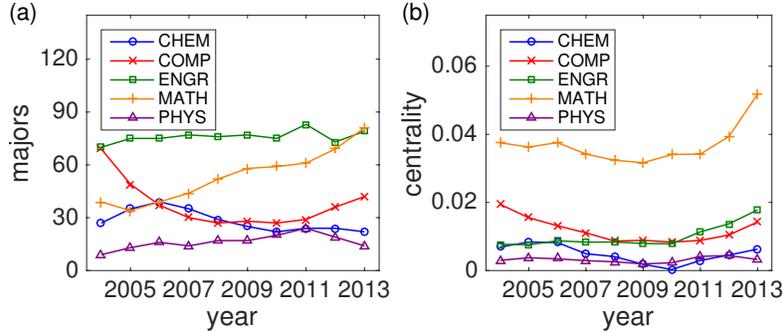, clip =,width=0.64\linewidth }
\caption{{\bf Eigenvector centrality: Quantitative sciences.} Evolution of the (a) number of majors and (b) eigenvector centrality of chemistry, computer science, engineering, mathematics, and physics from 2004-2013 using $\omega_{\text{cen}}=5$.} \label{fig4}
\end{figure}

Here we focus on classical eigenvector centrality of our network, using $C^{(t)}=A^{(t)}$, which not unlike PageRank centrality values nodes that are nearby other important nodes. The eigenvector centrality of a given node tends to be (but is not always) positively correlated with the degree of that node. We begin by investigating the centralities of the stalwart departments of the quantitative sciences community found above: chemistry, computer science, engineering, mathematics, and physics. For reference, we plot the number of majors present in each department during each year in Fig.~\ref{fig4}(a). (We forgo plotting their minors due to the fact that these particular departments do not offer disciplinary minors and therefore have little effect on the centralitles.) In Fig.~\ref{fig4}(b) we plot the evolution of each department's eigenvector centrality over the last ten years, computed using $\omega_{\text{cen}} = 5$ (and $\alpha=0.5$). First, we note that of these five departments engineering has on average the most majors, followed by mathematics, computer science, chemistry, then physics. However, mathematics has by far the largest centrality score. In hindsight we find that (i) a larger percentage of mathematics students also major or minor in another department and (ii) the other major or minor chosen by mathematics majors are surprisingly broad -- in addition to sharing majors and minors with the other quantitative sciences a significant number of mathematics students share majors or minors with department such as classical studies, economics, music, and philosophy (particularly in the most recent years).

\begin{figure}[t]
\centering
\epsfig{file =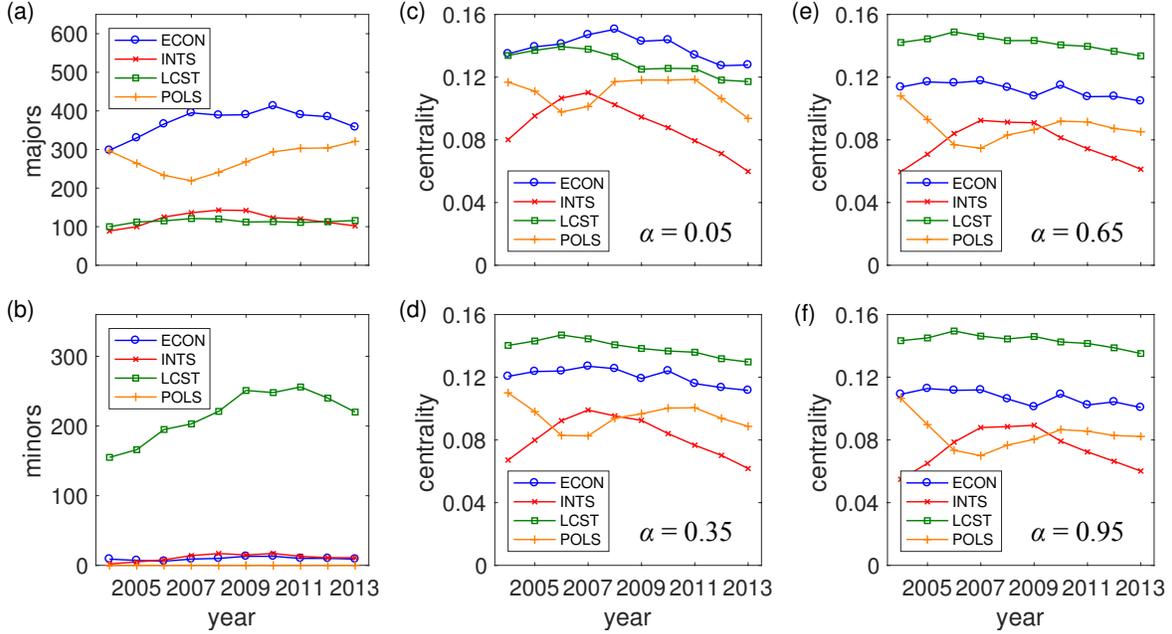, clip =,width=0.96\linewidth }
\caption{{\bf Eigenvector centrality: Effect of minors.} Evolution of the (a) number of majors and (b) number of minors of economics, international studies, language and culture studies, and political science from 2004-2013. For the same years, the eigenvector centralities of each department using $\omega_{\text{cen}}=5$ and minor parameters (c) $\alpha=0.05$, (d) $0.35$ (e) $\alpha=0.65$, and (d) $0.95$.} \label{fig5}
\end{figure}

We also use our network centrality measure to investigate the effect that minors have on the overall network structure. In particular, we study the eigenvector centrality of the four most central departments over the last ten years: economics, international studies, language and culture studies, and political science. In Figs.~\ref{fig5}(a) and (b) we plot the number of majors and minors, respectively, of these four popular departments. Economics has by far the most majors, followed by political science, then international studies, then language and culture studies. However language and culture studies has far more minors than the other three. (This is due to the fact that a large number of students complete a minor in a foreign language, all of which are housed in the language and culture studies department.) To highlight the role that minors play, we next compute the centralities for these four departments using a small minor parameter, $\alpha=0.05$, intermediate minor parameters, $\alpha=0.35$ and $0.65$, and a large minor parameter, $\alpha=0.95$, plotting the results in Figs.~\ref{fig5}(c) and (d). In the former case with $\alpha=0.05$ the evolution of the departments' centralities are reasonably well-described by the number of majors shown in Fig.~\ref{fig5}(a), except that language and culture studies is perhaps more central than expected, but still ranks below economics. However, as $\alpha$ is increased through $0.35$, $0.65$, and eventually to $0.95$, language and culture studies'  centrality is strengthened by its large number of minors, making it the most central department in the college by a significant margin.

\section*{Discussion}
The academic landscape of today's colleges and universities are organized by complex, time-varying networks that describe interactions between different departments~\cite{berger2002influence}. Moreover, the structural features of these academic networks have a strong impact on the activity of faculty members and the endeavors of students~\cite{pascarella2005college}. While social networks within colleges and universities have been studied in the past~\cite{bernard1980informant,guimera2003self}, the {\it academic} network structures describing interactions between departments is poorly understood. This leaves administration and individual departments to make decisions based on more simplistic statistical measure, without a robust understanding of the structure of the institution as a whole. To address this shortcoming, we have presented in this paper a framework constructing such an academic network and performed an analysis of community structure and centrality.

Our approach stems from the construction of a temporal multiplex network based on the double major, major-minor, and double minor patterns of graduating students. In particular, by representing departments as nodes and years as layers, we construct for each year a network based on the number of students that each pair of departments shares. Network features can then be extracted from any individual layer, or from the multiplex as a whole. Here we have focused on the two key features of community structure and centrality, using our home institution of Trinity College in Hartford, CT as an example. Beginning with community structure, we find that the community structure in any given year is more nuanced than the expected breakdown of sciences vs humanities. Rather, the sciences tends to break down into two communities, roughly corresponding to quantitative and descriptive sciences, while the humanities also tend to break down into two communities, roughly corresponding to historical and artistic humanities. Interestingly, recent years show a breakdown of, roughly speaking, historical and political humanities, artistic humanities, quantitative sciences, and descriptive sciences. However, through time these departments split and combine with one another, and certain departments switch between different communities while other stalwart departments remain in the same community. We also use time-varying eigenvector centrality to identify departments that are particularly important in connecting the college and study the effect that minors play in determining the relative standing of different departments.

These results have several practical applications. For instance, policy designed for the sciences will likely impact departments in different ways, depending on whether it is a quantitative or descriptive science. Thus, we hypothesize that in certain cases, taking the more subtle structure of the institution into account might result in more effective policies, and in other cases separate policies should be implemented targeting different parts of the college. Additionally, these results indicate an academic structure that might be more segregated than is ideal. To better unify the academic environment of the college, departments could focus on developing partnerships and interactions with other departments outside their community rather than inside. Moreover, we have used the time-evolution of the community structure to find stalwart departments that tend to remain in the same community, while others switch between communities intermittently, identifying which departments have more or less flexible interactions with their fellow departments.

Second, we have found that the evolution of the eigenvector centrality of a department reveals more than just the relative size of the department. We emphasize that a department's centrality does not necessarily correspond to its individual importance, but rather its importance in connecting the college as a whole. For instance, while mathematics has a moderate number of majors compared to the other quantitative sciences, it is highly central due to both its high number and diversity of double majors. Additionally, we can differentiate departments that are central due to their major influence (e.g., economics and political science) from those that are central due to their minor influence (e.g., language and culture studies).

While we have applied this approach to our home institution, we note that it is flexible and can in principle be applied to any other college or university where data describing the degrees of graduating seniors can be obtained. This opens the possibility for other researchers and institution officials to perform similar studies on their own college or university in order to better craft policies. A natural question then arises: how ``similar'' are the network structures at different colleges and universities? For instance: At different institutions, do communities break down into similar categories as we have found at Trinity? Do highly central departments at one institution tend to also be more central at other institutions? Are there significant differences in the structure of liberal arts colleges vs larger universities? We hypothesize that the framework presented here can be used to give insight into these questions.

\section*{Appendix}
\subsection*{Department codes}
The degrees awarded by Trinity College belong to 32 different departments. Here we identify each department with a different four-letter key, summarized in Table~\ref{tab1}.
\begin{table}[h]
\centering
\begin{tabular}{|l|l||l|l|}
\hline
Code & Department & Code & Department \\
\hline
AMST & American Studies & INTS & International Studies \\
ANTH & Anthropology & LCST & Language and Culture Studies \\
ARTH & Art History & MATH & Mathematics \\
BIOC & Biochemistry & MUSI & Music \\
BIOL & Biology & NEUR & Neuroscience \\
CHEM & Chemistry & PHIL & Philosophy \\
CLST & Classical Studies & PHYS & Physics \\
COMP & Computer Science & POLS & Political Science \\
ECON & Economics & PSYC & Psychology \\
EDUC & Educational Studies & PPLW & Public Policy and Law \\ 
ENGR & Engineering & RELI & Religion \\
ENGL & English & SOCI & Sociology \\
ENVI & Environmental Science & STUD & Studio Arts\\
FILM & Film Studies & THDA & Theatre and Dance \\
HIST & History & URBS & Urban Studies \\
INDP & Individualized Degree Program & WGSE & Women, Gender and Sexuality \\
\hline
\end{tabular}
\caption{Key of department codes.}\label{tab1}
\end{table}

\subsection*{Multiplex community detection}
As discussed in Ref.~\cite{bazzi2016community}, community detection in multiplex networks involves several subtle challenges. However, the optimization of multiplex modularity, i.e., Eq.~(\ref{eq:Q2}), can often be done using a modification of existing techniques for detecting communities in simple monoplex networks. Here we use a modification of the extremal optimization approach~\cite{duch2005community} summarized as follows. We begin by finding the community structures for each individual layer using extremal optimization. Next, we re-index the communities such that the Hamming distance between the communities for each pair of adjacent layers in minimized. (The Hamming distance between the communities of layers $t$ and $t+1$ is simply the number of nodes for which $s_i^{(t)}\ne s_i^{(t+1)}$.) At this point the community structures for each isolated layer have been found and are best-matched, maximizing the first term on the right hand-side of Eq.~(\ref{eq:Q2}). Finally, we sweep through each node in each layer in a random order, adjusting its membership to the community that locally optimizes Eq.~(\ref{eq:Q2}). Here we perform a total of 100 such sweeps. 

We note that finding community structure both in the isolated layers as well as in the layered multiplex includes a stochastic element. Therefore, the results presented in the main text represent the best outcome of 500 realization of maximizing the modularity in the multiplex. We find that the result for each realization is locally stable, i.e., changing the community membership of any one single node in a single layer decreases the overall multiplex modularity $Q_\omega$.

\section*{Acknowledgements}
The authors thank Terry Hosig in the Registrar's Office at Trinity College for help in obtaining the data. P.S.S. thanks Lauren Kiely Skardal for many helpful discussions. M.A. and S.W. acknowledge financial support from the Summer Student Research Program at Trinity College.

\bibliographystyle{plain}

\end{document}